\RequirePackage{amssymb}
\documentclass[sigconf,nonacm]{acmart}
\usepackage[OT2, T1]{fontenc}
\usepackage[russian, english]{babel}
\usepackage{amsmath,amsthm}
\usepackage{xcolor}
\usepackage{xspace}
\usepackage{setspace}
\usepackage{color,svgcolor}
\usepackage{paralist}
\usepackage{tikz}
\usepackage{multirow}
\usetikzlibrary{matrix}
\usepackage[capitalize]{cleveref}
\usepackage{array} 
\makeatletter
\def\@mathcolor#1#2#3{%
\protect\leavevmode
\begingroup
\color#1{#2}#3%
\endgroup
}
\def\mathcolor#1#{\@mathcolor{#1}}
\makeatother
\def\triadone{brown}
\def\triadtwo{red}
\def\triadthree{blue}
\def\triadfour{green}
\def\triadfive{purple}
\def\triadsix{dimgray}
\def\triadseven{chocolate}
\newtheorem{definition}{Definition}[section]
\newtheorem{definitions}{Definitions}[section]
\newtheorem{lemma}{Lemma}[section]

\newtheorem{remark}{Remark}[section]

\usepackage{thmtools}
\usepackage{thm-restate}

\newcommand{\norm}[1]{\left\lVert#1\right\rVert}
\newcommand{\matrixsize}[2]{\ensuremath{{{#1}\times{#2}}}}
\def\IdentityMatrix{\mathrm{Id}}
\newcommand{\IdentityMatrixOfSize}[1]{\ensuremath{{\IdentityMatrix}_{#1}}}
\def\tensorproduct{\otimes}
\def\tensor#1{{\mathcal{#1}}}
\newcommand{\MatrixRank}[1]{\ensuremath{{\textup{rank}\,{#1}}}}
\def\Trace{\textup{Trace}}

\def\MatrixProduct#1#2{{{#1}\cdot{#2}}}

\newcommand{\TensorRank}[3]{\ensuremath{{\textrm{R}\langle{#1}\times{#2}\times{#3}\rangle}}}
\newcommand{\FMMAF}[3]{\ensuremath{{\langle{#1}\times{#2}\times{#3}\rangle}}}
\newcommand{\FMMA}[4]{\ensuremath{{\langle{#1}\times{#2}\times{#3}:{#4}\rangle}}}
\newcommand{\FAMMA}[4]{\ensuremath{{\lbrace{#1}\times{#2}\times{#3}:{#4}\rbrace}}}
\newcommand{\MatrixSpace}[2]{\ensuremath{{{#1}^{#2}}}}

\newenvironment{smatrix}{\left(\begin{smallmatrix}}{\end{smallmatrix}\right)}
\newcommand{\Transpose}[1]{{{#1}^{\intercal}}\xspace}

\newcommand{\mathsc}[1]{{\normalfont\textsc{#1}}}

\newcommand{\Z}{\ensuremath{\mathbb{Z}}}

\newcommand{\F}{\ensuremath{\mathbb{F}}}

\newcommand{\BigO}[1]{\ensuremath{{O\mathopen{}({#1})\mathclose{}}}\xspace}

\newcommand{\Contraction}[2]{{\mathopen{}\left\{{#1},{#2}\right\}\mathclose{}}\xspace}

\makeatletter
\newcommand{\CWHILE}[2][default]{\algorithmicwhile\ #2\ %
\algorithmicdo\hfill%
\ALC@com{#1}\begin{ALC@whl}}
\newcommand{\CFORALL}[2][default]{\ALC@it\algorithmicforall\ #2\ %
\algorithmicdo\newline%
\ALC@com{#1}\begin{ALC@for}}
\newcommand{\CIF}[2][default]{\ALC@it\algorithmicif\ #2\ %
\algorithmicthen\hfill%
\ALC@com{#1}\begin{ALC@if}}
\newcommand{\FORALLDOEND}[2]{\ALC@it\algorithmicforall\ #1\ \algorithmicdo\ #2 \algorithmicendfor}
\makeatother
\usepackage[colorinlistoftodos]{todonotes}
\title{The tensor rank of~${{5}\times{5}}$ matrices multiplication \\ is bounded by~$98$ and its border rank by~$89$}
\author{Alexandre Sedoglavic}
\orcid{0000-0003-3225-8631}
\affiliation{
  \department{\textsc{umr} \textsc{cnrs} 9189 \textsc{cristal}}
  \institution{Universit\'e de Lille}
  \city{F-59000 Lille}
  \country{France}
}
\author{Alexey V.\ Smirnov}
\affiliation{%
  \department{Russian Federal Center of Forensic Science}
  \institution{Department of Justice}
  \city{Moscow}
  \country{Russia}
}
\makeatletter
\hypersetup{%
	pdftitle={\@title},
	pdfauthor={\@author},
	breaklinks=true,
	colorlinks=true,
	linkcolor=darkred,
	citecolor=blue,
	urlcolor=darkgreen,
}
\makeatother
\usepackage[num]{hypdestopt}
\begin{document}
\begin{abstract}
We present a non-commutative algorithm for the product of~${{3}\times{5}}$ by~${{5}\times{5}}$ matrices using~$58$ multiplications.
This algorithm allows to construct a non-commutative algorithm for multiplying~$\matrixsize{5}{5}$ (resp.~$\matrixsize{10}{10}, \matrixsize{15}{15}$) matrices using~$98$ (resp.~$686, 2088$) multiplications.
Furthermore, we describe an approximate algorithm that requires~$89$ multiplications and computes this product with an arbitrary small error.
\end{abstract}
\maketitle
\section{Introduction}\label{sec:Introduction}
Even if matrix multiplication is one of the most fundamental tool in scientific computing, it is still not completely understood.
Strassen's algorithm~\cite{strassen:1969}, with~$7$ recursive multiplications and~$18$ additions, was the first sub-cubic time algorithm for matrix product (with a cost of~$\BigO{n^{\log_{2}{7}}}$) and finding explicit algorithms for small size matrix product remains today a challenge.
In order to describe the contributions presented in this paper, let us recall some well-known terminologies for the sake of clarity.
\begin{definitions}\label{def:tensor}
We denote by~$\mathcal{M}_{m,n,p}$ the bilinear map representing the matrix product of a~$\matrixsize{m}{n}$-matrix by a~$\matrixsize{n}{p}$ matrix.
In particular, given any field~$\F$, there exist~$r$~$\matrixsize{m}{p}$ matrices~${{(C_{i})}_{{1}\leq{i}\leq{r}}}$,~${r}$ linear forms~${{(\ell_{i})}_{{1}\leq{i}\leq{r}}}$ from~$\MatrixSpace{\F}{\matrixsize{m}{n}}$ into~$\F$ and~${r}$ linear forms~${{(\ell^{\prime}_{i})}_{{1}\leq{i}\leq{r}}}$ from~$\MatrixSpace{\F}{\matrixsize{n}{p}}$ into~$\F$ such that the product of the~$\matrixsize{m}{n}$ matrix~$A$ by the~$\matrixsize{n}{p}$ matrix~$B$ is computed by the following computational scheme:
\begin{equation}
\mathcal{M}_{m,n,p}(A,B)
=\MatrixProduct{A}{B}=\sum_{i=1}^{r}\ell_{i}(A)\ell^{\prime}_{i}(B)C_{i}.
\end{equation}
This non-commutative scheme is classically interpreted as a \emph{tensor} (see precise encoding in Section~\ref{sec:UsedEncoding}) and we recall that the number~$r$ of its summands is the \emph{rank} of that tensor.
In this work, the notations~$\FMMA{m}{n}{p}{r}$ stands for a tensor of rank~$r$ encoding the product~$\mathcal{M}_{m,n,p}$.
We denotes by~$\FMMAF{m}{n}{p}$ the whole family of such schemes independently of their rank.
The \emph{tensor rank}~$\TensorRank{m}{n}{p}$ of the considered matrix product is the smallest integer~$r$ such that there is a tensor~$\FMMA{m}{n}{p}{r}$ in~$\FMMAF{m}{n}{p}$.
Similarly,~$\FAMMA{m}{n}{p}{r}$ denotes a computational scheme of rank~$r$ involving a parameter~$\epsilon$ whose limit computes the matrix product~$\mathcal{M}_{m,n,p}$ exactly as~$\epsilon$ tends to zero.
The \emph{border rank} of~$\mathcal{M}_{m,n,p}$ is the smallest integer~$r$ such that there exits an approximate scheme~$\FAMMA{m}{n}{p}{r}$.
\end{definitions}
The tensor rank~$\TensorRank{n}{n}{n}$ is related to the number of multiplications needed to compute the product of two~$\matrixsize{n}{n}$ and gives a measures of its complexity such as the exponent~$\omega$ of matrix multiplication equal to~${{\underline{\lim}}_{n\to\infty}\TensorRank{n}{n}{n}}$.
There is no result relative to asymptotic complexity in the present work and we refer to~\cite{Alman:2020aa} for a presentation of these researches, bibliographic references and the last best asymptotic bound~$\BigO{n^{2.3728596}}$ known to date.
\par
Alongside these galactic algorithms, a substantial amount of work was devoted to the practical design of computational schemes for small matrix products (see~\cite{drevet:2011a} for bibliographic references and~\cite{Sedoglavic:FMMDB} for last known results).
Our contribution follows this path and improves the complexity of the product of~$\matrixsize{5}{5}$ matrices and several other algorithms.
\par
The structure of~$\FMMAF{2}{2}{2}$ is broadly understood: 
there is basically just one rank~$7$ tensor~$\FMMA{2}{2}{2}{7}$ introduced in~\cite{strassen:1969} up to symmetries as shown in~\cite[Thm~0.1]{groot:1978}; 
furthermore, in that case the tensor rank is known to be~$7$ (see~\cite[Thm~3.1]{winograd:1971}).
But already the structure of~$\FMMAF{3}{3}{3}$ remains unresolved and very little is known about the actual complexity of small matrix products.
\par
Pursuing the works done the last~$53$ years in~\cite{schachtel:1978,makarov:1986b,makarov:1987,Sedoglavic:2017ab}, we propose a new non-commutative algorithm for multiplying~${{5}\times{5}}$ matrices using~$98$ multiplications.
More precisely, by presenting explicit algorithms, we establish the following result:
\begin{theorem}\label{thm:main}
The tensor rank of matrix multiplication tensor encoding the product of a~$\matrixsize{3}{5}$ by a~$\matrixsize{5}{5}$ matrix is bounded by~$58$.
\par
This implies that the tensor rank of matrix multiplication tensor encoding the product of~$\matrixsize{5}{5}$ matrices is bounded by~$98$.
\par
Furthermore, the border rank of matrix multiplication tensor encoding the product of~$\matrixsize{5}{5}$ matrices is bounded by~$89$.
\end{theorem}
This theorem results mainly from the work initiated by the second author in~\cite{smirnov:2013a}.
This work induces many algorithms (e.g.\ see~\cite{Sedoglavic:FMMDB} for a complete list) and especially an algorithm~$\FMMA{3}{3}{6}{40}$ that leads to tensors~$\FMMA{6}{6}{6}{160}$ and~$\FMMA{12}{12}{12}{1040}$ (this last one defines a generic algorithm with cost~$\BigO{n^{2.796}}$ for suitable square matrices).
\section{Tensor representation of matrix multiplication}\label{sec:UsedEncoding}
In the forthcoming sections, we present several matrix product algorithms as trilinear forms.
To do so, let us review the needed notions through an well-known example.
The matrix product~${C=\MatrixProduct{A}{B}}$ could be computed using Strassen algorithm by performing the following computations (see~\cite{strassen:1969}):
\begin{equation}\label{eq:StrassenMultiplicationAlgorithm}
\begin{array}{ll}
\mathcolor{\triadone}{\rho_{1}}\leftarrow{\mathcolor{\triadone}{a_{11}}(\mathcolor{\triadone}{b_{12}-b_{22}})},
&
\\
\mathcolor{\triadtwo}{\rho_{2}}\leftarrow{(\mathcolor{\triadtwo}{a_{11}+a_{12}})\mathcolor{\triadtwo}{b_{22}}},
&
\mathcolor{\triadfour}{\rho_{4}}\leftarrow{(\mathcolor{\triadfour}{a_{12}-a_{22}})(\mathcolor{\triadfour}{b_{21}+b_{22}})},
\\
\mathcolor{\triadthree}{\rho_{3}}\leftarrow{(\mathcolor{\triadthree}{a_{21}+a_{22}}) \mathcolor{\triadthree}{b_{11}}},
&
\mathcolor{\triadfive}{\rho_{5}}\leftarrow{(\mathcolor{\triadfive}{a_{11}+a_{22}})(\mathcolor{\triadfive}{b_{11}+b_{22}})},
\\
\mathcolor{\triadsix}{\rho_{6}}\leftarrow{\mathcolor{\triadsix}{a_{22}}(\mathcolor{\triadsix}{b_{21}-b_{11}})},
&
\mathcolor{\triadseven}{\rho_{7}}\leftarrow{(\mathcolor{\triadseven}{a_{21}-a_{11}})(\mathcolor{\triadseven}{b_{11}+b_{12}})},
\\[\medskipamount]
\multicolumn{2}{c}{
\begin{smatrix} c_{11} &c_{12} \\ c_{21} &c_{22} \end{smatrix}
=
\begin{smatrix}
\mathcolor{\triadfive}{\rho_{5}} + \mathcolor{\triadfour}{\rho_{4}} - \mathcolor{\triadtwo}{\rho_{2}} + \mathcolor{\triadsix}{\rho_{6}} &
\mathcolor{\triadsix}{\rho_{6}} + \mathcolor{\triadthree}{\rho_{3}} \\
\mathcolor{\triadtwo}{\rho_{2}} + \mathcolor{\triadone}{\rho_{1}}&
\mathcolor{\triadfive}{\rho_{5}} + \mathcolor{\triadseven}{\rho_{7}} + \mathcolor{\triadone}{\rho_{1}}- \mathcolor{\triadthree}{\rho_{3}}
\end{smatrix}\!.}
\end{array}
\end{equation}
In order to consider this algorithm under a geometric standpoint, we present it as a tensor.
Matrix multiplication is a bilinear map:
\begin{equation}
\begin{array}{ccl}
\F^{\matrixsize{m}{n}} \times \F^{\matrixsize{n}{p}} & \rightarrow &\F^{\matrixsize{m}{p}}, \\
(A,B) &\rightarrow & \MatrixProduct{A}{B},
\end{array}
\end{equation}
where the spaces~$\F^{\matrixsize{\cdot}{\cdot}}$ are finite vector spaces over a field~$\F$ that can be endowed with the Frobenius inner product:
\begin{equation}{{\langle M,N\rangle}={\Trace({\MatrixProduct{\Transpose{M}}{N}})}}.\end{equation}
Hence, this inner product establishes an isomorphism between~$\F^{\matrixsize{\cdot}{\cdot}}$ and its dual space~${\bigl(\F^{\matrixsize{\cdot}{\cdot}}\bigr)}^{\star}$ allowing for example to associate matrix multiplication and the trilinear form~${\Trace(\MatrixProduct{\Transpose{C}}{\MatrixProduct{A}{B}})}$:
\begin{equation}\label{eq:TrilinearForm}
\begin{array}{ccc}
\F^{\matrixsize{m}{n}} \times \F^{\matrixsize{n}{p}} \times {(\F^{\matrixsize{m}{p}})}^{\star}&\rightarrow & \F, \\
(A,B,\Transpose{C}) &\rightarrow & \langle {C},\MatrixProduct{A}{B}\rangle.
\end{array}
\end{equation}
As by construction, the space of trilinear forms is the canonical dual space of order three tensor product, we could associate the Strassen multiplication algorithm~(\ref{eq:StrassenMultiplicationAlgorithm}) with the tensor~$\tensor{S}$ defined by:
\begin{equation}\label{eq:StrassenTensor}
\begin{array}{r}
\sum_{i=1}^{7}{P_{i}}\!\tensorproduct\!{Q_{i}}\!\tensorproduct\!{S_{i}}=
\mathcolor{\triadone}{\begin{smatrix}1&0\\0&0\\\end{smatrix}\!\tensorproduct\!\begin{smatrix}0&1\\0&-1\\\end{smatrix}\!\tensorproduct\!\begin{smatrix}0&0\\1&1\\\end{smatrix}
}
\!+\!\\[\bigskipamount]
\mathcolor{\triadtwo}{\begin{smatrix}1&1\\0&0\\\end{smatrix}\!\tensorproduct\!\begin{smatrix}0&0\\0&1\\\end{smatrix}\!\tensorproduct\!\begin{smatrix}-1&0\\1&0\\\end{smatrix}}
\!+\!
\mathcolor{\triadthree}{\begin{smatrix}0&0\\1&1\\\end{smatrix}\!\tensorproduct\!\begin{smatrix}1&0\\0&0\\\end{smatrix}\!\tensorproduct\!\begin{smatrix}0&1\\0&-1\end{smatrix}}
\!+\!\\[\bigskipamount]
\mathcolor{\triadfour}{\begin{smatrix}0&1\\0&-1\\\end{smatrix}\!\tensorproduct\!\begin{smatrix}0&0\\1&1\\\end{smatrix}\!\tensorproduct\!\begin{smatrix}1&0\\0&0\\\end{smatrix}}
\!+\!
\mathcolor{\triadfive}{{\begin{smatrix}1&0\\0&1\end{smatrix}}\!\tensorproduct\!{\begin{smatrix}1&0\\0&1\end{smatrix}}\!\tensorproduct\!\begin{smatrix}1&0\\0&1\\\end{smatrix}}
\!+\!\\[\bigskipamount]
\mathcolor{\triadsix}{\begin{smatrix}0&0\\0&1\\\end{smatrix}\!\tensorproduct\!\begin{smatrix}-1&0\\1&0\\\end{smatrix}\!\tensorproduct\!\begin{smatrix}1&1\\0&0\\\end{smatrix}}
\!+\!
\mathcolor{\triadseven}{\begin{smatrix}-1&0\\1&0\\\end{smatrix}\!\tensorproduct\!\begin{smatrix}1&1\\0&0\\\end{smatrix}\!\tensorproduct\!\begin{smatrix}0&0\\0&1\\\end{smatrix}}
\!
\end{array}
\end{equation}
in~${{(\F^{\matrixsize{m}{n}})}^{\star} \tensorproduct {(\F^{\matrixsize{n}{p}})}^{\star} \tensorproduct \F^{\matrixsize{m}{p}}}$ with~${m=n=p=2}$.
As stated in Definitions~\ref{def:tensor}, this tensor is a particular description of the matrix product bilinear map.
Indeed, there is an infinite set of \emph{equivalent} tensors shown by the following theorem:
\begin{theorem}[{\cite[\S~2.8]{groot:1978a}}]\label{thm:isotropygroup}
The isotropy group of any matrix multiplication tensor in~$\FMMAF{m}{n}{p}$ is~${{({\mathsc{psl}(m)}\times{\mathsc{psl}(n)}\times{\mathsc{psl}(p)})}\rtimes{\mathfrak{S}_{3}}}$, where~$\mathsc{psl}$ stands for the group of matrices of determinant~${\pm{1}}$ and~$\mathfrak{S}_{3}$ for the symmetric group on~$3$ elements.
\end{theorem}
The following proposition gives another description from an invariant perspective.
\begin{proposition}[{\cite[\S~2.5.1]{Landsberg:2016ab}}]\label{prop:MatrixProductIntrinsicDefinition}
The tensor defining the product of a~$\matrixsize{m}{n}$-matrix by a~$\matrixsize{n}{p}$-matrix is isomorphic to the tensor~${{\IdentityMatrixOfSize{\matrixsize{m}{m}}}\tensorproduct{\IdentityMatrixOfSize{\matrixsize{n}{n}}}\tensorproduct{\IdentityMatrixOfSize{\matrixsize{p}{p}}}}$.
This isomorphism gives the well-known expression of the classical matrix multiplication tensor:
\begin{equation}\label{eq:ClassicalMatrixMultiplicationTensor}
\sum_{i=1}^{m} \sum_{j=1}^{n} \sum_{k=1}^{p} {E_{i}^{j}}\tensorproduct{E_{j}^{k}}\tensorproduct{E_{k}^{i}},
 \end{equation}
where~${E_{i}^{j}}$ denotes the matrix with its coefficient at the intersection of line~$i$ and column~$j$ equal to~$1$ and all its other coefficients equal to~$0$.
\end{proposition}
Let us introduce now an invariant for the action described in Theorem~\ref{thm:isotropygroup} that will be useful in our presentation.
\begin{definition}\label{def:type}
Given a tensor~$\tensor{P}$ decomposable as sum of rank-one elementary tensors:
\begin{equation}
\label{eq:5}
\tensor{P}=\sum_{i=1}^{q} {P_{i}}\tensorproduct{Q_{i}}\tensorproduct{S_{i}}
\end{equation}
where~$P_{i}, Q_{i}$ and~$S_{i}$ are matrices of suitable sizes for~$i$ in~${\{1,\ldots,q\}}$.
\par
The \emph{type} of a tensor~$\tensor{P}$ is the list~${{[{(\MatrixRank{P_{i}},\MatrixRank{Q_{i}},\MatrixRank{S_{i}})}]}_{i={{1}\ldots{q}}}}$.
\par
Inspired by~\cite[\S~4]{Heule:2019ab}, we encode the type of a tensor~$\tensor{P}$ as the following polynomial:
\begin{equation}\label{eq:TypePolynomialRepresentation}
{T_{\tensor{P}}(X,Y,Z)}={\sum_{i=1}^{q} X^{\MatrixRank{P_{i}}}Y^{\MatrixRank{Q_{i}}}Z^{\MatrixRank{S_{i}}}}.
\end{equation}
\end{definition}
Hence, the type of Strassen's tensor is~${{X}^{2}{Y}^{2}{Z}^{2}+6\,XYZ}$.
\par
Now we precise the trilinear form of tensor that is used in the sequel of this work.
Given any triple~${(A,B,C)}$ of suitable size matrices, one can explicitly express from tensor~$\tensor{S}$ the Strassen matrix multiplication algorithm computing~$\MatrixProduct{A}{B}$ by the \emph{complete} contraction~${\Contraction{\tensor{S}}{{A}\tensorproduct{B}\tensorproduct{C}}}$:
\begin{equation}\label{eq:TensorAction}
\begin{array}{c}
	\left({(\MatrixSpace{\F}{\matrixsize{m}{n}})}^{\star}\! \tensorproduct \! {(\F^{\matrixsize{n}{p}})}^{\star}\! \tensorproduct\! \F^{\matrixsize{m}{p}} \right)\! \tensorproduct \! \left(
\F^{\matrixsize{m}{n}}\! \tensorproduct\! \F^{\matrixsize{n}{p}} \tensorproduct{{(\F^{\matrixsize{m}{p}})}^{\star}}\right)
\!\rightarrow\!
\F, \\[\medskipamount]
\tensor{S}\tensorproduct{({A}\tensorproduct{B}\tensorproduct{C})} \rightarrow
\sum_{i=1}^{7} \langle {P_{i}}, A \rangle \langle {Q_{i}}, B \rangle \langle S_{i}, C\rangle
\end{array}
\end{equation}
that is equal to~${\Trace(\MatrixProduct{\MatrixProduct{A}{B}}{C})}$.
Hence, for inputs~${A={(a_{ij})},B={(b_{ij})}}$ and~${C=(c_{ij})}$ of suitable sizes~(${i=1,2}$ and~${j=1,2}$), the representation of Strassen Algorithm~(\ref{eq:StrassenMultiplicationAlgorithm}) as a trilinear form is:
\begin{equation}
\begin{array}{c}
{a_{11}}\left({b_{12}}-{b_{22}}\right)\left({c_{21}}+{c_{22}}\right)\\
+\left({a_{11}}+{a_{12}}\right){b_{22}}\left(-{c_{11}}+{c_{21}}\right)\\
+\left({a_{21}}+{a_{22}}\right){b_{11}}\left({c_{12}}-{c_{22}}\right)\\
+\left({a_{12}}-{a_{22}}\right)\left({b_{21}}+{b_{22}}\right){c_{11}}\\
+\left({a_{11}}+{a_{22}}\right)\left({b_{11}}+{b_{22}}\right)\left({c_{11}}+{c_{22}}\right)\\
+{a_{22}}\left({b_{21}}-{b_{11}}\right)\left({c_{11}}+{c_{12}}\right)\\
+\left({a_{21}}-{a_{11}}\right)\left({b_{11}}+{b_{12}}\right){c_{22}}.
\end{array}
\end{equation}
Before ending this section let us recall that, as stated in introduction, the~$\FMMAF{2}{2}{2}$ is pretty well-understood even in its rectangular counterpart as shown by the following proposition:
\begin{proposition}[{\cite[Thm~1]{Hopcroft:1971}}]\label{prop:Hopcroft2xmxn}
The~$\matrixsize{m}{2}$ by~$\matrixsize{2}{n}$ matrix product can be encoded by a~$\FMMA{m}{2}{n}{\lceil(3mn+\max(m,n))/2\rceil}$ tensor.
\end{proposition}
The next section is devoted to gather classical results used in the sequel of this paper to construct new algorithm for small matrix product.
\subsection{Constructing composite tensors from small ``atomic'' ones}
Introducing tensor to represents matrix product and their relationship with the trace operator induces naturally several interesting results on matrix product algorithms.
First, let us remark that---given three matrices~${A,B}$ and~$C$ of suitable sizes---the following properties of the trace operator:
\begin{equation}\label{eq:12.1}
\begin{array}{lll}
    {\Trace\left(A\cdot B\cdot C\right)} &={\Trace\left(C\cdot A\cdot
        B\right)},\\ &= {\Trace\left(B\cdot C\cdot A\right)}, \\
    &= \Trace\big( \Transpose{\!\left(A\cdot B\cdot C\right)}\big), \\ & =
    \Trace\big( \Transpose{C} \cdot\Transpose{B}\cdot\Transpose{A}
    \big),
\end{array}
\end{equation}
show that the following relations hold:
\begin{equation}
\begin{array}{lll}
\FMMA{m}{n}{p}{r}&=\FMMA{p}{m}{n}{r}&=\FMMA{n}{p}{m}{r}, \\
&=\FMMA{p}{n}{m}{r}&=\FMMA{m}{p}{n}{r},\\
&=\FMMA{n}{m}{p}{r}.
\end{array}
\end{equation}
Furthermore, for all~$\ell$ such that~${{1}\leq{\ell}\leq{m-1}}$ there is a natural isomorphism between~${{\MatrixSpace{\F}{\matrixsize{m}{n}}}}$ and~${{{\MatrixSpace{\F}{\matrixsize{\ell}{n}}}\oplus{\MatrixSpace{\F}{\matrixsize{(m-\ell)}{n}}}}}$; 
The same remark shows that~$\IdentityMatrixOfSize{\matrixsize{m}{m}}$ is equal to~${\IdentityMatrixOfSize{\matrixsize{\ell}{\ell}}\oplus\IdentityMatrixOfSize{\matrixsize{(m-\ell)}{(m-\ell)}}}$.
This relation and the tensor product's properties imply that the tensor~${{{\IdentityMatrixOfSize{\matrixsize{m}{m}}} \tensorproduct {\IdentityMatrixOfSize{\matrixsize{n}{n}}} \tensorproduct {\IdentityMatrixOfSize{\matrixsize{p}{p}}}}}$ is equal to
\begin{equation}
{{\IdentityMatrixOfSize{\matrixsize{\ell}{\ell}}}\tensorproduct{\IdentityMatrixOfSize{\matrixsize{n}{n}}}\tensorproduct{\IdentityMatrixOfSize{\matrixsize{p}{p}}}}+{{\IdentityMatrixOfSize{\matrixsize{(m-\ell)}{(m-\ell)}}}\tensorproduct{\IdentityMatrixOfSize{\matrixsize{n}{n}}}\tensorproduct{\IdentityMatrixOfSize{\matrixsize{p}{p}}}}.
\end{equation}
These simple remarks and Proposition~\ref{prop:MatrixProductIntrinsicDefinition} recall that small sizes matrix product algorithms allow by their direct sum to construct an algorithm for the product of matrices of bigger sizes as shown by the following well-known lemma:
\begin{lemma}\label{lem:AddCombining}
Given~$\FMMA{\ell}{n}{p}{r}$ and~$\FMMA{(m-\ell)}{n}{p}{s}$, one can construct~$\FMMA{m}{n}{p}{r+s}$ as follow:
\begin{equation}
\FMMA{m}{n}{p}{r+s}=\FMMA{\ell}{n}{p}{r}\oplus\FMMA{(m-\ell)}{n}{p}{s}.
\end{equation}
There is a similar construction using the tensor product:
\begin{equation}
\FMMA{mu}{nv}{pw}{rs}={\FMMA{m}{n}{p}{r}}\tensorproduct{\FMMA{u}{v}{w}{s}}.
\end{equation}
\end{lemma}
We say that a tensor is ``atomic'' if it is not constructed using the above constructions, Proposition~\ref{prop:Hopcroft2xmxn} or if it is induced by Strassen's algorithm
(see~\cite[\S~2]{drevet:2011a} and~\cite{Sedoglavic:2017ac} for such constructions).
\par
The previous sections were devoted to the notions and notations necessary to describe concisely the new results on which the next sections focus.
We could now present the main result of this paper.
\section{New exact algorithms}\label{sec:3x5x5_58}
The rank of the classical tensor in~$\FMMAF{3}{5}{5}$ is~$75$.
Proposition~\ref{prop:Hopcroft2xmxn} and Lemma~\ref{lem:AddCombining} improves the resulting bound to~$63$ as shown by the following relations:
\begin{align}
\FMMA{3}{3}{5}{38}=\FMMA{3}{3}{3}{23}\oplus\FMMA{3}{3}{2}{15},\\
\FMMA{3}{5}{5}{63}=\FMMA{3}{3}{5}{38}\oplus\FMMA{3}{2}{5}{25}.
\end{align}
This bound was superseded using~$\FMMA{3}{3}{5}{36}$ introduced in~\cite{smirnov:2013a} and above standard constructions.
\par
Our results are also rooted in the same kind of experimental mathematics combining computer power and human efforts.
Thus, let us give a short account of a method allowing to find new atomic tensors in the next section.
\subsection{How to construct small rank matrix multiplication tensors}\label{sec:Method}
As any tensor in~$\FMMAF{m}{n}{p}$ encodes the same bilinear map, Proposition~\ref{eq:ClassicalMatrixMultiplicationTensor} implies that the following relation always holds:
\begin{equation}\label{eq:BrentSystem}
\FMMA{m}{n}{p}{r} - \sum_{i=1}^{m} \sum_{j=1}^{n} \sum_{k=1}^{p} {E_{i}^{j}}\tensorproduct{E_{j}^{k}}\tensorproduct{E_{k}^{i}} = 0.
\end{equation}
Using an ansatz with undetermined coefficients for~$\FMMA{m}{n}{p}{r}$, this relation defines the Brent over-determined system of~${(mnp)}^{2}$ cubic polynomial equations in~$({mn+np+pm})r$ unknowns (see~\cite[\S~5, eq~5.03]{brent:1970a}).
Theoretically, a Gr{\"o}bner basis computation allows to describe all solutions of this system and thus close the topic.
But such a resolution is not possible in practice for any matrix size of interest.
Nevertheless several original matrix multiplication algorithms where found by hand (e.g.~$\FMMA{3}{3}{3}{23}$ in~\cite{laderman:1976a} and probably Strassen's algorithm~\cite{strassen:1969}) and almost every method for solving were tried (e.g.\ \textsc{sat} solver in~\cite[\S~2]{Heule:2019ab}) with---up to our knowledge---few complexity improvements.
\par
For now, the almost only productive approach remains numerical optimization method using least-squares methods and heuristics.
In fact, while objective functions derived from Equation~(\ref{eq:BrentSystem}) are non-convex and nonlinear, they can be splitted in three linear subsystems (by helding two components of the unknown tensor fixed for example) and their resolution boils down to linear algebra.
Nevertheless, in order to obtain new results, a regularization term needs to be added and the following is chosen in this paper:
\begin{equation}
\!\!\!\!\!
\begin{array}{c}
\textrm{Arg}\ \min\\
{P_{i},Q_{i},S_{i}}
\end{array}\!\!\!\!\!\!
\begin{array}{r}
\norm{\sum_{i=0}^{q}
{P_{i}}\!\tensorproduct\!{Q_{i}}\!\tensorproduct\!{S_{i}}
-\sum_{i=0}^{m} \sum_{j=0}^{n} \sum_{k=0}^{p}
{E_{i}^{j}}\!\tensorproduct\!{E_{j}^{k}}\!\tensorproduct\!{E_{k}^{i}} 
}\\[\medskipamount]
+
\lambda\left(\sum_{i=0}^{q}
\norm{P_{i}-\widetilde{P_{i}}}+
\norm{Q_{i}-\widetilde{Q_{i}}}+
\norm{S_{i}-\widetilde{S_{i}}}
\right).
\end{array}
\end{equation}
with \emph{model} matrices~${\widetilde{P_{i}},\widetilde{Q_{i}},\widetilde{S_{i}}}$ defining the regularization and a scalar parameter~$\lambda$ that determines the weight of the regularization term.
The models are designed to drive the solution to match a desired structure and are choosed carefully for each iteration (see~\cite{smirnov:2013a} for a detailed presentation).
This approach gives a uniform method for deriving exact algorithm but also approximate one when they are a order~$1$ polynomial approximation w.r.t\ their parameter~$\epsilon$.
\par
A better precision---an approximation order greater then~$1$---requires several other heuristics. 
If all dimensions of the problem are greater than~$3$, it is not yet possible to obtain acceptably short exact algorithms. 
Hence, the success of the resolution presented here relies on heuristical expertise and tyazhelaya rabota.
However, the works ot that topic done since~\cite{smirnov:2013a} show that the objective function of the found approximate algorithms allows to presumably estimate the exact rank of large problems.
This point will appear in a future work.
\par
Let us now describe the latest exact matrix multiplication tensor found with this method.
\subsection{$\FMMA{3}{5}{5}{58}$ description}\label{seq:3x5x5:58_Description}
Before the detailed description done in the next section, let us first present the type introduced in Definition~\ref{def:type} of this tensor:
\begin{equation}
\begin{array}{c}
17\,{X}^{2}{Y}^{2}{Z}^{2}+2\,X{Y}^{4}Z+{X}^{3}{Y}^{2}Z+X{Y}^{2}{Z}^{3}\\
+5\,{X}^{3}YZ+5\,XY{Z}^{3}+2\,{X}^{2}{Y}^{2}Z+2\,X{Y}^{2}{Z}^{2}+X{Y}^{3}Z\\
+{X}^{2}YZ+XY{Z}^{2}+13\,X{Y}^{2}Z\\
+7\,XYZ=T(X,Y,Z).
\end{array}
\end{equation}
\begin{remark}
Considering that the indeterminates commute, the relation~${T(X,Y,Z)=T(Z,Y,X)}$ holds.
Unfortunately, even if this property suggests the existence of a symmetry (see Equation~(\ref{eq:12.1}) and~\cite[$\pi_{13}$ in Thm~3.4]{groot:1978a} for a detailed description), this tensor does not have a non-trivial stabilizer.
Such stabilizer are not uncommon for tensor found using the method sketched in Section~\ref{sec:Method}: for example the stabilizer~${{({(C_{2}\times{C_{2}})}\rtimes{S_{4}})}\rtimes{C_{2}}}$ of~$\FMMA{3}{3}{6}{40}$ is of order~$192$ (see~\cite{Sedoglavic:FMMDB}).
So, even if Comon's conjecture was disproved in full generality (see~\cite{Shitov:2017ab}), there might be a tensor in~$\FMMAF{3}{5}{5}$ of rank~$58$ with stabilizer~${{{C_{2}\times{C_{2}}}\times{S_{3}}}}$.
\end{remark}
As two matrix multiplication tensor of same rank could share the same type, this invariant is not a faithful descriptions and we give its explicit expression in the next section.
\paragraph{Trilinear form of~$\FMMA{3}{5}{5}{58}$.}
We split the forthcoming description in~$13$ expressions whose components have the type corresponding to their indices as follow:
\begin{equation}\begin{split}
\FMMA{3}{5}{5}{58}&=\tau_{222}+\tau_{141}+\tau_{321}+\tau_{123}+\tau_{311}\\
&+\tau_{113}+\tau_{221}+\tau_{122}+\tau_{131}+\tau_{211}\\
&+\tau_{112}+\tau_{121}+\tau_{111}.
\end{split}\end{equation}
Let us start with~$\tau_{141}$:
\begin{equation}\begin{split}
\tau_{141}=&
{a_{24}}\left(\begin{aligned}&{b_{13}}+{b_{22}}+{b_{23}}-{b_{24}}\\&+{b_{33}}-{b_{35}}-{b_{43}}+{b_{44}}+{b_{45}}\end{aligned}\right)\left({c_{41}}+{c_{42}}\right)\\
-&\left({a_{14}}-{a_{24}}\right)
\left(\begin{aligned}&{b_{14}}-{b_{32}}+{b_{34}}+{b_{41}}\\&+{b_{42}}-{b_{44}}+{b_{51}}-{b_{54}}-{b_{55}}\end{aligned}\right){c_{41}}.
\end{split}\end{equation}
\begin{equation}\begin{split}
\tau_{211}=\left({a_{12}}-{a_{22}}-{a_{25}}\right)\left({b_{22}}-{b_{24}}\right)\left({c_{12}}+{c_{32}}+{c_{42}}\right)\\
\end{split}\end{equation}
\begin{equation}\begin{split}
\tau_{112}=\left({a_{12}}+{a_{14}}-{a_{15}}\right){b_{41}}\left({c_{11}}+{c_{31}}+{c_{12}}\right).
\end{split}\end{equation}
\begin{equation}\begin{split}
\tau_{222}=&\left({a_{12}}-{a_{22}}+{a_{23}}\right)\left({b_{22}}+{b_{31}}\right)\left({c_{11}}+{c_{12}}+{c_{21}}+{c_{41}}\right)\\
+&\left({a_{32}}-{a_{22}}+{a_{23}}\right)\left({b_{22}}-{b_{33}}\right)\left({c_{23}}+{c_{43}}-{c_{32}}-{c_{33}}\right)\\
+&\left({a_{33}}+{a_{15}}+{a_{35}}\right)\left({b_{52}}-{b_{31}}\right)\left({c_{11}}+{c_{21}}+{c_{41}}-{c_{13}}\right)\\
+&\left({a_{13}}+{a_{14}}+{a_{34}}\right)\left({b_{42}}+{b_{33}}+{b_{35}}\right)
	\left(
\begin{aligned}
&{c_{31}}+{c_{23}}\\&-{c_{33}}+{c_{43}}
\end{aligned}
\right)\\
+&\left({a_{25}}-{a_{31}}-{a_{35}}\right)\left({b_{14}}+{b_{55}}\right)\left({c_{52}}-{c_{42}}-{c_{43}}\right)\\
+&\left({a_{24}}-{a_{11}}-{a_{14}}\right)\left({b_{14}}-{b_{35}}+{b_{45}}\right)\left({c_{52}}-{c_{41}}-{c_{42}}\right)\\
+&\left({a_{21}}+{a_{12}}-{a_{22}}\right)\left({b_{11}}+{b_{25}}\right)\left({c_{11}}+{c_{51}}+{c_{12}}\right)\\
+&\left({a_{21}}-{a_{22}}+{a_{32}}\right)\left({b_{13}}-{b_{25}}\right)\left({c_{32}}+{c_{33}}-{c_{53}}\right)\\
+&\left({a_{31}}+{a_{15}}+{a_{35}}\right)\left({b_{55}}-{b_{11}}\right)\left({c_{11}}+{c_{51}}-{c_{13}}\right)\\
+&\left({a_{12}}+{a_{24}}\right)\left(
\begin{aligned}
&{b_{43}}+{b_{24}}-{b_{22}}\\&-{b_{13}}-{b_{23}}-{b_{33}}
\end{aligned}
\right)
	\left({c_{41}}+{c_{32}}+{c_{42}}\right)\\
+&\left({a_{11}}+{a_{14}}+{a_{34}}\right)\left({b_{13}}-{b_{35}}+{b_{45}}\right)\left({c_{31}}-{c_{33}}+{c_{53}}\right)\\
+&\left(
\begin{aligned}
&{a_{13}}+{a_{14}}-{a_{21}}\\&-{a_{23}}-{a_{24}}
\end{aligned}
\right)\left({b_{12}}+{b_{35}}\right)\left({c_{51}}-{c_{21}}-{c_{22}}\right)\\
+&\left({a_{24}}-{a_{14}}+{a_{15}}\right)\left(-{b_{41}}-{b_{51}}+{b_{54}}+{b_{55}}\right)\left({c_{41}}-{c_{12}}\right)\\
+&\left({a_{34}}+{a_{15}}\right)\left({b_{41}}+{b_{53}}\right)\left({c_{31}}+{c_{13}}\right)\\
+&\left({a_{32}}+{a_{25}}\right)\left({b_{22}}-{b_{24}}+{b_{53}}\right)\left({c_{32}}-{c_{43}}\right)\\
+&\left({a_{33}}-{a_{25}}+{a_{35}}\right)\left({b_{12}}-{b_{32}}+{b_{52}}+{b_{34}}\right)\left({c_{43}}-{c_{22}}\right)\\
+&\left({a_{13}}+{a_{14}}-{a_{24}}\right)\left({b_{34}}+{b_{42}}-{b_{32}}\right)\left({c_{41}}-{c_{22}}\right).
\end{split}\end{equation}
\begin{equation}\begin{split}
\tau_{321}=&
\left(\begin{aligned}&{a_{13}}-{a_{23}}+{a_{33}}+{a_{14}}\\&-{a_{24}}-{a_{25}}+{a_{35}}\end{aligned}\right)\left({b_{12}}-{b_{32}}+{b_{34}}\right){c_{22}}.
\end{split}\end{equation}
\begin{equation}\begin{split}
\tau_{123}=&
-\left({a_{13}}+{a_{14}}\right)\left({b_{42}}+{b_{35}}\right)\left(\begin{aligned}&{c_{31}}-{c_{21}}-{c_{22}}\\&+{c_{23}}-{c_{33}}+{c_{43}}\end{aligned}\right).
\end{split}\end{equation}
\begin{equation}\begin{split}
\tau_{311}=
&\left(\begin{aligned}&{a_{13}}+{a_{15}}+{a_{22}}+{a_{33}}\\&+{a_{35}}-{a_{12}}-{a_{23}}\end{aligned}\right){b_{31}}\left({{c_{21}}+c_{11}}+{c_{41}}\right)\\
+&\left(\begin{aligned}&{a_{23}}+{a_{32}}-{a_{13}}-{a_{22}}\\&-{a_{14}}-{a_{33}}-{a_{34}}\end{aligned}\right){b_{33}}\left({c_{23}}-{c_{33}}+{c_{43}}\right)\\
+&\left(\begin{aligned}&{a_{11}}-{a_{21}}+{a_{31}}+{a_{14}}\\&-{a_{24}}-{a_{25}}+{a_{35}}\end{aligned}\right){b_{14}}\left({c_{52}}-{c_{42}}\right)\\
+&\left(\begin{aligned}&{a_{11}}-{a_{21}}+{a_{31}}-{a_{12}}\\&+{a_{22}}+{a_{15}}+{a_{35}}\end{aligned}\right){b_{11}}\left({c_{11}}+{c_{51}}\right)\\
+&\left(\begin{aligned}&{a_{21}}+{a_{32}}-{a_{11}}-{a_{14}}\\&-{a_{22}}-{a_{31}}-{a_{34}}\end{aligned}\right){b_{13}}\left({c_{53}-{c_{33}}}\right).
\end{split}\end{equation}
\begin{equation}\begin{split}
\tau_{113}=& 
\left({a_{22}}-{a_{23}}\right){b_{22}}\left(\begin{aligned}&{c_{11}}+{c_{12}}+{c_{21}}+{c_{22}}+{c_{23}}\\&-{c_{32}}-{c_{33}}+{c_{41}}+{c_{42}}+{c_{43}}\end{aligned}\right)\\
-&\left({a_{31}}+{a_{35}}\right){b_{55}}\left(\begin{aligned}&{c_{11}}-{c_{13}}+{c_{42}}+{c_{43}}\\&+{c_{51}}-{c_{52}}-{c_{53}}\end{aligned}\right)\\
+&\left({a_{11}}+{a_{14}}\right)\left({b_{35}}-{b_{45}}\right)\left(\begin{aligned}&{c_{31}}-{c_{33}}+{c_{41}}+{c_{42}}\\&-{c_{51}}-{c_{52}}+{c_{53}}\end{aligned}\right)\\
+&\left({a_{22}}-{a_{21}}\right){b_{25}}\left(\begin{aligned}&{c_{11}}+{c_{12}}-{c_{32}}\\-&{c_{33}}+{c_{51}}+{c_{52}}+{c_{53}}\end{aligned}\right)\\
-&\left({a_{33}}+{a_{35}}\right){b_{52}}\left({c_{11}}+{c_{21}}+{c_{41}}-{c_{22}}-{c_{13}}-{c_{23}}\right).
\end{split}\end{equation}
\begin{equation}\begin{split}
\tau_{221}=&\left({a_{22}}-{a_{32}}+{a_{24}}\right)\left({b_{22}}+{b_{13}}+{b_{23}}+{b_{33}}-{b_{24}}\right){c_{32}}\\
-&\left({a_{12}}+{a_{14}}+{a_{34}}\right)\left({b_{41}}+{b_{13}}+{b_{33}}-{b_{43}}\right){c_{31}}.
\end{split}\end{equation}
\begin{equation}\begin{split}
\tau_{122}={a_{15}}\left({b_{41}}+{b_{51}}-{b_{52}}-{b_{55}}\right)\left({c_{11}}+{c_{41}}-{c_{13}}\right)\\
+{a_{25}}\left({b_{22}}-{b_{24}}+{b_{54}}+{b_{55}}\right)\left({c_{12}}+{c_{42}}+{c_{43}}\right).
\end{split}\end{equation}
\begin{equation}
\tau_{131}=\left({a_{25}}-{a_{35}}\right)\left(\begin{aligned}&{b_{12}}+{b_{14}}-{b_{32}}+{b_{34}}\\&+{b_{52}}+{b_{53}}-{b_{54}}\end{aligned}\right){c_{43}}.
\end{equation}
\begin{equation}\begin{split}
\tau_{121}=&{a_{32}}\left({b_{21}}-{b_{53}}\right){c_{13}}\\
-&{a_{12}}\left(-{b_{21}}+{b_{41}}+{b_{22}}+{b_{25}}\right)\left({c_{11}}+{c_{12}}\right)\\
+&{a_{32}}\left({b_{22}}+{b_{23}}-{b_{53}}+{b_{25}}\right)\left({c_{32}}+{c_{33}}\right)\\
+&{a_{34}}\left({b_{42}}+{b_{43}}+{b_{53}}+{b_{45}}\right)\left({c_{33}}-{c_{31}}\right)\\
+&{a_{13}}\left({b_{32}}-{b_{31}}-{b_{42}}\right)\left({c_{21}}+{c_{41}}\right)\\
+&{a_{33}}\left({b_{32}}+{b_{33}}+{b_{35}}-{b_{12}}-{b_{52}}\right)\left({c_{23}}+{c_{43}}\right)\\
+&{a_{23}}\left({b_{22}}+{b_{34}}\right)\left({c_{22}}+{c_{42}}\right)\\
+&{a_{31}}\left({b_{13}}+{b_{15}}-{b_{55}}\right){c_{53}}\\
+&{a_{21}}\left({b_{14}}+{b_{15}}+{b_{25}}-{b_{35}}\right){c_{52}}\\
+&{a_{11}}\left({b_{12}}+{b_{15}}+{b_{35}}-{b_{11}}-{b_{45}}\right){c_{51}}\\
+&{a_{34}}\left({b_{44}}+{b_{53}}-{b_{42}}\right){c_{43}}\\
-&\left({a_{15}}+{a_{35}}\right)\left({b_{11}}+{b_{31}}-{b_{51}}+{b_{53}}\right){c_{13}}\\ 
+&\left({a_{22}}-{a_{12}}\right)\left({b_{11}}+{b_{21}}+{b_{31}}+{b_{22}}-{b_{24}}\right){c_{12}}.
\end{split}\end{equation}
\begin{equation}\begin{split}\label{eq:Tau111}
\tau_{111}=&\left({a_{31}}+{a_{33}}\right){b_{12}}{c_{23}}\\
+&\left({a_{21}}+{a_{23}}+{a_{24}}-{a_{11}}-{a_{13}}-{a_{14}}\right){b_{12}}\left({c_{51}}-{c_{21}}\right)\\
+&\left({a_{21}}+{a_{23}}+{a_{24}}\right){b_{35}}\left({c_{51}}+{c_{52}}-{c_{21}}-{c_{22}}\right)\\
+&\left({a_{33}}+{a_{34}}\right){b_{35}}\left({c_{53}}-{c_{23}}-{c_{43}}\right)\\
+&{a_{12}}\left({b_{13}}+{b_{23}}+{b_{33}}-{b_{43}}\right)\left({c_{31}}+{c_{41}}+{c_{32}}+{c_{42}}\right)\\
+&\left({a_{24}}+{a_{15}}-{a_{14}}-{a_{25}}\right)\left({b_{54}}+{b_{55}}-{b_{51}}\right){c_{12}}\\
+&\left({a_{32}}-{a_{34}}+{a_{35}}\right){b_{53}}\left({c_{13}}+{c_{33}}+{c_{43}}\right).
\end{split}\end{equation}
\paragraph{Induced geometry}
As shown by Theorem~\ref{thm:isotropygroup}, the action of the group~${{({\mathsc{psl}(3)}\times{\mathsc{psl}(5)}\times{\mathsc{psl}(5)})}\rtimes{{C}_{2}}}$ on the tensor introduced in the previous section defines classically a manifold of dimension~$52$ of tensors~$\FMMA{3}{5}{5}{58}$.
Furthermore, this tensor have~$8$ \emph{serendipitous equalities} that is couple of summands that shares the same factor (e.g.\ the two first summands of Equation~\ref{eq:Tau111} share the factor~$b_{12}$).
Up to our knowledge, this property was first introduced in~\cite[\S~9.3]{Smith:2002aa} but does not seem to receive the attention it deserves.
For example, this property allows to define new transformation of a matrix multiplication tensor into another as shown by the following lemma:
\begin{lemma}
Given any invertible~$\matrixsize{q}{q}$-matrix~$M$, the tensor with serendipitous equalities~${\sum_{i=0}^{q}{{U_{i}}\tensorproduct{V_{i}}\tensorproduct{W}}}$ involving the component~$W$ is equal to the tensor~${\sum_{i=0}^{q}{{\alpha_{i}}\tensorproduct{\beta_{i}}\tensorproduct{W}}}$ defined by: 
\begin{equation}
\begin{smatrix}\alpha_{1}\\\vdots\\\alpha_{q}\end{smatrix}
=\mathrm{Transpose}(M)\begin{smatrix}U_{1}\\\vdots\\U_{q}\end{smatrix},\quad
\begin{smatrix}\beta_{1}\\\vdots\\\beta_{q}\end{smatrix}
=M^{-1}\begin{smatrix}V_{1}\\\vdots\\V_{q}\end{smatrix}.
\end{equation}
\end{lemma}
The proof of this lemma reduces to the trivial computation of the expression~${{\sum_{i=0}^{q}{{U_{i}}\tensorproduct{V_{i}}\tensorproduct{W}}}-{{\alpha_{i}}\tensorproduct{\beta_{i}}\tensorproduct{W}}}$.
\par
This lemma shows that the dimension of the manifold induced by the tensor introduced in this section is greater then that could be expected.
\par
The following sections are devoted to present other interesting consequences.
\subsection{Upper bound~$98$ on tensor rank of~$\FMMAF{5}{5}{5}$}
Lemma~\ref{lem:AddCombining} and the tensor presented in Section~\ref{seq:3x5x5:58_Description} allows to construct the following tensor:
\begin{equation}\label{sec:5x5x5:98}
\FMMA{5}{5}{5}{98}=\FMMA{2}{5}{5}{40}\oplus\FMMA{3}{5}{5}{58},
\end{equation}
with the construction of tensor~$\FMMA{2}{5}{5}{40}$ taken from~\cite{Hopcroft:1971} (its explicit expression is given in~\cite{Sedoglavic:FMMDB}).
Remark that the best theoretical lower bound on the corresponding tensor rank is~$48$ (see~\cite[Theorem~2]{Blaser:1999aa}).
\par
Furthermore, this new atomic tensor also improves the construction of the following algorithms:
\begin{align}
{\FMMA{10}{10}{10}{686}}&=&{{\FMMA{2}{2}{2}{7}}\otimes{\FMMA{5}{5}{5}{98}}},\\
{\FMMA{15}{15}{15}{2088}}&=&{{\FMMA{3}{5}{5}{58}}\otimes{\FMMA{5}{3}{3}{36}}}.
\end{align}
All these tensors are explicitly presented via~\cite{Sedoglavic:FMMDB}.
\paragraph{Consequence of this new upper bound.}
A group-theoretic approach to the conception of matrix multiplication algorithm related to Fourier transform on finite groups was introduced by Cohn and Umans in~\cite{cohn:2003a}.
The new tensor constructed in Equation~(\ref{sec:5x5x5:98}) allows to exhibit a limitation of this approach as shown by the following remark.
\begin{remark}\label{rem:grouptheoretic}
It is shown in~\cite{hart:2013a} that no group can realize~$\matrixsize{5}{5}$ matrix multiplication better then Makarov's algorithm~$\FMMA{5}{5}{5}{100}$ using the group-theoretic approach of Cohn and Umans~\cite{cohn:2003a}.
Hence, the tensor presented in this note shows that this approach does not produce better algorithms then~$\FMMA{5}{5}{5}{98}$.
The same assertion holds for~$\FMMA{3}{3}{3}{23}$ and~$\FMMA{4}{4}{4}{49}$ (see~\cite[Theorem~7.3]{Hedtke:2012aa}).
\end{remark}
The next section is devoted to describe a new approximate algorithm~${\FAMMA{5}{5}{5}{89}}$.
\section{New approximate algorithms}
Approximate matrix multiplication tensors were first introduced by Bini et ali in~\cite{Bini:1979} in order to improve asymptotic bounds. 
From a practical point of view, these approximate algorithms could be used efficiently when the coefficients are in~${\Z/p\Z}$ (see~\cite{Boyer:2014aa}).
Furthermore, from a theoretical point of view, these tensors allow to work with Euclidean closure of the Brent algebraic variety defined by Equation~(\ref{eq:BrentSystem}) and not the Zariski closure induced by dealing with exact tensors. 
As the Zariski closure is often much larger then the Euclidean closure, this shift of standpoint brings usually lower bounds.
Hence, the exact tensor~$\FMMA{2}{3}{3}{15}$ presented in~\cite{Hopcroft:1971} is optimal; Smirnov describes~$\FAMMA{2}{3}{3}{14}$ in~\cite{smirnov:2013a} and that bound on the corresponding border rank was proved to be optimal in~\cite[Theorem~1.4]{Conner:2019ac}.
Similarly, while the best upper bound for tensor rank of~$\FMMAF{3}{3}{3}$ is~$23$ (\cite{laderman:1976a}),~$\FAMMA{3}{3}{3}{20}$ could be found in~\cite{smirnov:2013a} (\cite[Theorem~1.1]{Conner:2019ac} reports that the lower bound is~$17$ for the corresponding border rank).
\par
Lemma~\ref{lem:AddCombining} shows that these atomic approximate matrix multiplication tensors lead to the following tensor used in the sequel:
\begin{equation}\label{eq:approx_5x3x3}
{\FAMMA{5}{3}{3}{34}}={{\FAMMA{3}{3}{3}{20}}\oplus{\FAMMA{2}{3}{3}{14}}}.
\end{equation}
The results in~\cite{Bini:1979} are based on a \emph{partial} matrix multiplication algorithm that computes approximately the product of a~$\matrixsize{2}{2}$-matrix~$A$ with one element vanishing~(e.g.~${a_{22}=0}$) and a~$\matrixsize{2}{2}$-matrix~$B$ a full matrix.
This kind of tensor were used to improve the bound on the exponent of matrix multiplication (see~\cite[\S~3]{schonhage:1981}).
\par
Let us now show how to complete a tensor constructed using Equation~\ref{eq:approx_5x3x3} in order to define~$\FAMMA{5}{5}{5}{89}$.
\subsection{A partial approximate matrix multiplication}
This section presents a partial approximate tensor~$\tensor{T}_{\epsilon}$ that defines an algorithm computing the product of a~$\matrixsize{5}{5}$-matrix~$A$ with~$9$ vanishing elements:
\begin{equation}
a_{ij}=0,\ \forall (i,j)\ \textrm{such that}\ {1}\leq{i}\leq{3},\ {3}\leq{j}\leq{5}.
\end{equation}
and a full~$\matrixsize{5}{5}$-matrix~$B$.
\paragraph{The type of~$\tensor{T}_{\epsilon}$}
Remarks that the isotropy introduced in Theorem~\ref{thm:isotropygroup} and the associated the invariant presented in Definition~\ref{def:type} for exact matrix multiplication tensors remain obviously valid for approximates ones.
The type of~$\tensor{T}_{\epsilon}$ is:
\begin{equation}\label{eq:PartialApproximateMatrixMultiplicationType}
\begin{array}{c}
20\,{X}^{2}{Y}^{2}{Z}^{2}\\
+3\,{X}^{2}{Y}^{2}Z+2\,{X}^{2}Y{Z}^{2}+4\,X{Y}^{2}{Z}^{2}\\
+7\,{X}^{2}YZ+6\,X{Y}^{2}Z+8\,XY{Z}^{2}\\
+5\,XYZ.
\end{array}
\end{equation}
Again, we are going to split its description by trilinear form into several summands whose subscript indicate the type of their components as follow:
\begin{equation}
\tensor{T}_{\epsilon}=\rho_{222}+\rho_{221}+\rho_{212}+\rho_{122}+\rho_{211}+\rho_{121}+\rho_{112}+\rho_{111}.
\end{equation}
\begin{equation}\begin{split}\label{eq:Approx222}
\rho_{222}=&
\left(a_{11}-a_{22}\epsilon^{3}\right)\left(b_{12}-b_{21}-b_{11}\epsilon^{3}\right)\left(c_{21}+\frac{c_{12}}{\epsilon^{3}}\right)\\
+&\left(a_{11}+a_{52}\epsilon^3\right)\left(\begin{aligned}&b_{41}+b_{51}-b_{31}\\&+\frac{b_{15}}{\epsilon^{2}}+\frac{b_{21}-b_{13}}{\epsilon^{3}}\end{aligned}\right)\left({c_{15}}+c_{51}{\epsilon^{2}}\right)\\
+&\left(a_{11}+a_{53}+a_{52}\epsilon^3\right)\left(b_{13}+b_{31}\epsilon^{3}\right)\left(c_{31}+\frac{c_{51}}{\epsilon}+\frac{c_{15}}{\epsilon^{3}}\right)\\
+&\left(a_{41}-a_{43}+\frac{a_{55}}{\epsilon^{3}}\right)\left({b_{34}}+b_{53}\epsilon^{3}\right)\left(c_{35}-c_{44}-c_{45}\epsilon^{3}\right)\\
+&\left(\begin{aligned}&a_{11}-a_{54}\\&+a_{51}\epsilon+a_{52}\epsilon^{3}\end{aligned}\right)\left({b_{14}}-b_{41}\epsilon^{3}\right)\left(c_{41}+\frac{c_{15}}{\epsilon^{3}}\right)\\
+&\left(a_{21}+a_{54}\right)\left(b_{42}-\frac{b_{45}}{\epsilon}+\frac{b_{14}}{\epsilon^{3}}\right)\left(c_{25}+c_{42}\epsilon^{3}\right)\\
+&\left(a_{21}-a_{53}\right)\left({b_{13}} +b_{35}\epsilon^{2} -b_{32}\epsilon^{3} \right)\left(c_{32}+\frac{c_{52}}{\epsilon}+\frac{c_{25}}{\epsilon^{3}}\right)\\
+&\left(a_{22}+a_{45}\right)\left(b_{55}+\frac{b_{23}}{\epsilon^{3}}\right)\left({c_{54}}+c_{32}{\epsilon^{3}}\right)\\
+&\left(a_{32}+a_{44}\right)\left(b_{41}+\frac{b_{24}}{\epsilon^{3}}\right)\left({c_{14}}+c_{43}{\epsilon^{3}}\right)\\
+&\left(a_{12}+a_{43}\right)\left(b_{32}+\frac{b_{25}}{\epsilon^{3}}\right)\left({c_{24}}+c_{51}{\epsilon^{3}}\right)\\
+&\left(a_{22}+a_{41}\right)\left(b_{15}+\frac{b_{22}}{\epsilon^{3}}\right)\left({c_{54}}+c_{22}{\epsilon^{3}}\right)\\
+&\left(a_{43}-a_{52}\right)\left(b_{33}+\frac{b_{25}}{\epsilon^{3}}\right)\left({c_{34}}-c_{55}{\epsilon^{3}}\right)\\
+&\left(a_{45}-a_{32}\right)\left(b_{51}+\frac{b_{23}}{\epsilon^{3}}\right)\left({c_{14}}-c_{33}{\epsilon^{3}}\right)\\
+&\left(a_{41}-a_{32}\right)\left(b_{11}+\frac{b_{22}}{\epsilon^{3}}\right)\left({c_{14}}-c_{23}{\epsilon^{3}}\right)\\
+&\left(a_{22}+a_{44}\right)\left(b_{45}-\frac{b_{24}}{\epsilon^{3}}\right)\left({c_{54}}-c_{42}{\epsilon^{3}}\right)\\
+&\left(a_{41}+a_{52}\right)\left(b_{13}-\frac{b_{22}}{\epsilon^{3}}\right)\left({c_{34}}-c_{25}{\epsilon^{3}}\right)\\
+&\left(a_{12}+a_{44}\right)\left(b_{42}-\frac{b_{24}}{\epsilon^{3}}\right)\left({c_{24}}-c_{41}{\epsilon^{3}}\right)\\
+&\left(a_{32}+a_{43}\right)\left(b_{31}-\frac{b_{25}}{\epsilon^{3}}\right)\left({c_{14}}-c_{53}{\epsilon^{3}}\right)\\
+&\left(a_{44}-a_{52}\right)\left(b_{43}-\frac{b_{24}}{\epsilon^{3}}\right)\left({c_{34}}+c_{45}{\epsilon^{3}}\right)\\
+&\left(a_{45}-a_{12}\right)\left(b_{52}-\frac{b_{23}}{\epsilon^{3}}\right)\left({c_{24}}+c_{31}{\epsilon^{3}}\right).\\
\end{split}\end{equation}
\begin{equation}\begin{split}\label{eq:Approx212}
\rho_{212}=&
\left(\frac{a_{51}}{{\epsilon}^{2}}-\frac{a_{31}+a_{53}}{\epsilon^{3}}\right)b_{13}\left({c_{15}}+c_{55}{\epsilon}-c_{53}\epsilon^{2}-c_{33}\epsilon^{3}\right)\\
+&\left(a_{31}-a_{54}\right)b_{14}\left(c_{43}-\frac{c_{15}}{{\epsilon}^{3}}-\frac{c_{55}}{\epsilon^{2}}\right).
\end{split}\end{equation}
\begin{equation}\begin{split}\label{eq:Approx221}
\rho_{221}=&\left(\frac{a_{55}-a_{11}}{\epsilon^{3}}-\frac{a_{51}}{\epsilon^{2}}-a_{52}\epsilon\right)\left({b_{14}}+b_{15}{\epsilon}+b_{51}{\epsilon}^{3}\right)c_{15}\\
+&\left(a_{41}+\frac{a_{55}}{\epsilon^{3}}\right)\left({b_{14}+b_{34}-b_{54}}+{b_{15}}\epsilon+b_{55}{{\epsilon}^{2}}\right)c_{44}\\
-&\left(\frac{a_{21}+a_{55}}{\epsilon^{3}}\right)\left(b_{14}+b_{15}{\epsilon}+b_{55}\epsilon^{2}-b_{52}\epsilon^{3}\right)c_{25}.
\end{split}\end{equation}
\begin{equation}\begin{split}\label{eq:Approx121}
\rho_{121}=&
{a_{32}}
\left(\frac{b_{21}+b_{22}+b_{23}-b_{24}+b_{25}}{\epsilon^{3}}-b_{41}\right)c_{14}\\
+&\left(a_{51}+\frac{a_{54}}{\epsilon^{3}}\right)\left({b_{14}}-b_{45}\epsilon^{2}+b_{44}\epsilon^{3}\right)c_{45}\\
+&\left(a_{51}+\frac{a_{53}}{\epsilon^{3}}\right)\left({b_{13}+b_{34}}+b_{35}\epsilon^{2}+b_{33}\epsilon^{3}\right)c_{35}\\
+&a_{52}\!\left(b_{33}+b_{43}-b_{13}-b_{53}+\frac{b_{22}+b_{23}-b_{24}+b_{25}}{\epsilon^{3}}\!\right)\!c_{34}\\
-&a_{12}\!\left(b_{32}+b_{42}-b_{12}-b_{52}+\frac{b_{22}+b_{23}-b_{24}+b_{25}}{\epsilon^{3}}\!\right)\!c_{24}\\
-&a_{22}\!\left(b_{15} +b_{45}+b_{55} -b_{35}+\frac{b_{22}+b_{23}-b_{24}+b_{25}}{\epsilon^{3}}\!\right)\!c_{54}.
\end{split}\end{equation}
\begin{equation}\begin{split}\label{eq:Approx122}
\rho_{122}=& 
a_{53}\left({b_{13}}+b_{35}{\epsilon^{2}}\right)
\left(
\begin{aligned}
&c_{32}-c_{33}-c_{31}+\frac{c_{52}-c_{53}-c_{51}}{\epsilon}\\
&+\frac{c_{55}}{\epsilon^{2}}+\frac{c_{25}-c_{35}}{\epsilon^{3}}
\end{aligned}
\right)\\
+&a_{21}{\epsilon^{2}}\left(
\begin{aligned}
&b_{32}-b_{42}-b_{52}+\frac{b_{15}}{\epsilon^{2}} \\
&+\frac{b_{55}+b_{45}-b_{35}}{\epsilon} +\frac{b_{12}-b_{13}}{\epsilon^{3}}
\end{aligned}
\right)\left(c_{52}+\frac{c_{25}}{\epsilon^{2}}\right)\\
+&\frac{a_{54}}{\epsilon^{2}}
\left(b_{45}-\frac{b_{14}}{\epsilon^{2}}\right)\left(
\begin{aligned}
&c_{42}-c_{41}-c_{43} \\
&+\frac{c_{55}}{\epsilon^{2}} +\frac{c_{25}+c_{45}}{\epsilon^{3}}
\end{aligned}
\right)\\
+&\frac{a_{55}}{\epsilon}\left(\frac{b_{14}}{\epsilon}+b_{15}+b_{55}\epsilon\right)\left( \frac{c_{25}-c_{44}}{\epsilon} +c_{55}\right).
\end{split}\end{equation}
\begin{equation}\begin{split}\label{eq:Approx211}
\rho_{211}=&
\left(a_{42}+\frac{a_{22}-a_{43}}{\epsilon^{3}}\right)\left({b_{25}}-b_{35}\epsilon^{3}\right)\left(c_{54}+c_{52}\epsilon^3\right)\\
+&\left(a_{45}+a_{52}-a_{42}\epsilon^{3}\right)\left({b_{23}}-b_{53}\epsilon^{3}\right)\left(c_{35}-\frac{c_{34}}{\epsilon^{3}}\right)\\
+&\left(a_{43}-a_{41}-a_{51}-\frac{a_{53}+a_{55}}{\epsilon^{3}}\right)b_{34}\left( c_{35} -c_{45}\epsilon^{3} \right)\\
+&\left(a_{41}+a_{45}+\frac{a_{55}}{\epsilon^{3}}\right)\left(b_{54}+b_{53}\epsilon^{3}\right)\left(c_{44}+c_{45}\epsilon^3\right)\\
+&\left(a_{42}+\frac{a_{12}-a_{41}}{\epsilon^{3}}\right)\left({b_{22}}-b_{12}\epsilon^{3}\right)\left(c_{24}+c_{21}\epsilon^3\right)\\
+&\left(\frac{a_{31}-a_{55}}{\epsilon}\right)\left(\frac{b_{14}}{\epsilon}+b_{15}\right)\left(\frac{c_{15}}{\epsilon}+c_{55}\right)\\
+&\left(a_{51}+a_{22}-\frac{a_{11}+a_{21}}{\epsilon^{3}}\right)\left({b_{12}}-b_{11}\epsilon^{3}\right)c_{12}.
\end{split}\end{equation}
\begin{equation}\begin{split}\label{eq:Approx112}
\rho_{112}=&
a_{43}b_{25}\left(c_{52}+c_{55}-c_{51}-c_{53}+\frac{c_{14}+c_{54}-c_{24}-c_{34}}{\epsilon^{3}}\right)\\
-&a_{41}b_{22}\left(c_{22}+c_{25}-c_{21}-c_{23}+\frac{c_{14}+c_{54}-c_{24}-c_{34}}{\epsilon^{3}}\right)\\
-&a_{45}b_{23}\left(c_{32}+c_{35}-c_{31}-c_{33}+\frac{c_{14}+c_{54}-c_{24}-c_{34}}{\epsilon^{3}}\right)\\
-&a_{44}b_{24}\left(c_{43}+\frac{c_{14}-c_{24}-c_{34}-c_{44}-c_{54}}{\epsilon^{3}}\right)\\
-&{a_{11}}b_{21}\left(\frac{c_{51}}{\epsilon}+\frac{c_{11}+c_{15}-c_{12}}{\epsilon^{3}}\right)\\
+&a_{31}b_{11}\left(c_{13}+c_{55}-c_{53}\epsilon+\frac{c_{15}}{\epsilon}\right)\\
+&\left({a_{21}}-a_{51}\epsilon^{3}\right)b_{12}\left(c_{22}-\frac{c_{52}}{\epsilon}-\frac{c_{25}-c_{12}}{\epsilon^{3}}\right)\\
+&\left(a_{51}-\frac{a_{31}}{\epsilon}\right)\left(b_{15}+b_{11}\epsilon-\frac{b_{13}}{\epsilon}\right)\left(c_{55}-c_{53}\epsilon+\frac{c_{15}}{\epsilon}\right).
\end{split}\end{equation}
\begin{equation}\begin{split}\label{eq:Approx111}
\rho_{111}=&\left(a_{42}-\frac{a_{44}}{{\epsilon}^{3}}\right)\!\left({b_{24}}-b_{44}\epsilon^{3}\right)\!\left((c_{41}+c_{42}-c_{45})\epsilon^{3}+{c_{44}}\right)\\
+&\left({a_{12}}{\epsilon}^{3}+a_{11}\right)\left(b_{11}\epsilon^{3}+\frac{b_{21}}{\epsilon}\right)\left(\frac{c_{11}}{\epsilon^{3}}+c_{21}\right)\\
+&\left(a_{32}-a_{42}\epsilon^{3}\right)\!\left(b_{11}-b_{31}+b_{51}-\frac{b_{21}}{\epsilon^{3}}\right)\!\left(c_{14}-c_{13}\epsilon^{3}\right)\\
+&a_{54}b_{43}c_{35} +a_{31}b_{12}c_{23}.\\
\end{split}\end{equation}
Remark that there is~$4$ serendipitous equalities in this tensor.
\par
We conclude the construction of an approximate tensor~$\FAMMA{5}{5}{5}{89}$ in the next section.
\subsection{New upper bound~$89$ on border rank of~$\FMMAF{5}{5}{5}$}
Using the approximate matrix multiplication tensor defined in the previous section and the construction made in Equation~(\ref{eq:approx_5x3x3}), one can construct easily:
\begin{equation}
\FAMMA{5}{5}{5}{89}=\tensor{T}_{\epsilon}+\FAMMA{3}{3}{3}{20} + \FAMMA{2}{3}{3}{14}.
\end{equation}
Remark that the best theoretical lower bound on the corresponding border rank is~$45$ (see~\cite[Corollary~1.2]{Landsberg:2015aa})).
\section{Perspectives}
We have presented there the upper bound~$98$ (resp.~$89$) for the tensor (resp.\ border) rank of the~$\matrixsize{5}{5}$-matrix product while the best theoretical lower bound is~$48$ (resp.~$45$) (see~\cite[Theorem~2]{Blaser:1999aa} (resp.~\cite[Corollary~1.2]{Landsberg:2015aa})).
\par
Furthermore, the new $\FMMA{3}{5}{5}{59}$ improves also
\begin{itemize}
\item ${{\FMMA{10}{10}{10}{686}}={{\FMMA{2}{2}{2}{7}}\otimes{\FMMA{5}{5}{5}{98}}}}$
\item ${{\FMMA{15}{15}{15}{2088}}={{\FMMA{3}{5}{5}{58}}\otimes{\FMMA{5}{3}{3}{36}}}}$
\end{itemize}
and we have a quite clear idea of all exact matrix multiplication tensors up to size~${{32}\times{32}\times{32}}$ induced by the best known-to-date ``atomic'' such tensors.
But as~${\FMMA{15}{15}{15}{2088}}$'s tensor rank is lesser then the first decomposition~${{\FMMA{3}{3}{3}{23}}\otimes{\FMMA{5}{5}{5}{98}}}$ that comes to mind, the most obvious decomposition does not always lead to the current most efficient one.
Hence, there is a---simple but requiring expensive calculation---computer based search to do in order to construct a database for approximate tensor similar to~\cite{Sedoglavic:FMMDB}.
\bibliographystyle{abbrvurl}
\bibliography{3x5x5_58}
\end{document}